\documentclass[12pt]{iopart}
\expandafter\let\csname equation*\endcsname\relax
  \expandafter\let\csname endequation*\endcsname\relax
\usepackage{latexsym,amsmath}
\usepackage{amsbsy}
\usepackage[pdftex]{graphicx}
\usepackage{amssymb}
\usepackage{epstopdf}
\usepackage[usenames]{color}
\usepackage{color}

\begin{document}
\title{Cosmological networks}

\author{Mari{\'a}n Bogu{\~n}{\'a}$^1$, Maksim Kitsak$^2$, and Dmitri Krioukov$^3$}
\address{$^1$ Departament de F{\'\i}sica Fonamental, Universitat de Barcelona,
Mart\'{\i} i Franqu\`es 1, 08028 Barcelona, Spain}
\address{$^2$ Center for Complex Network Research and Department of Physics, Northeastern University,
360 Huntington Avenue, Boston, MA 02115, USA}
\address{$^3$ Department of Physics, Northeastern University,
360 Huntington Avenue, Boston, MA 02115, USA}
\ead{marian.boguna@ub.edu}

\date{\today}

\begin{abstract}

Networks often represent systems that do not have a long history of studies in traditional fields of physics, albeit there are some notable exceptions such as energy landscapes and quantum gravity. Here we consider networks that naturally arise in cosmology. Nodes in these networks are stationary observers uniformly distributed in an expanding open FLRW universe with any scale factor, and two observers are connected if one can causally influence the other. We show that these networks are growing Lorentz-invariant graphs with power-law distributions of node degrees.
These networks encode maximum information about the observable universe available to a given observer.

\end{abstract}


\maketitle

\section{Introduction}

Network science is intrinsically multidisciplinary because the systems it studies come
from different domains of science. Complex networks are everywhere indeed---in
communication technologies, social and political sciences, biology and medicine,
economics, or even linguistics~\cite{DorMen-book03,newman03c-review,BoLaMoChHw06}. That
is why many fields of science---computer science, social sciences, biology, statistics,
mathematics, and certainly physics---have contributed tremendously over the last decade
to network research. Surprisingly, even though statistical physics has been applied with
great success to understanding complex networks, the systems that these networks
represent can rarely display a long history of broad interest and focused research in
traditional fields of physics. In fact, none of the network examples above provide an
exception to this general rule. Exceptions, such as energy landscape
networks~\cite{Doye2002} and networks in background-independent approaches to quantum
gravity~\cite{BoLe87,KrKi12,KoMa08,RoSp10}, are rare indeed.

\begin{figure}[t]
\centerline{\includegraphics[width=0.8\linewidth]{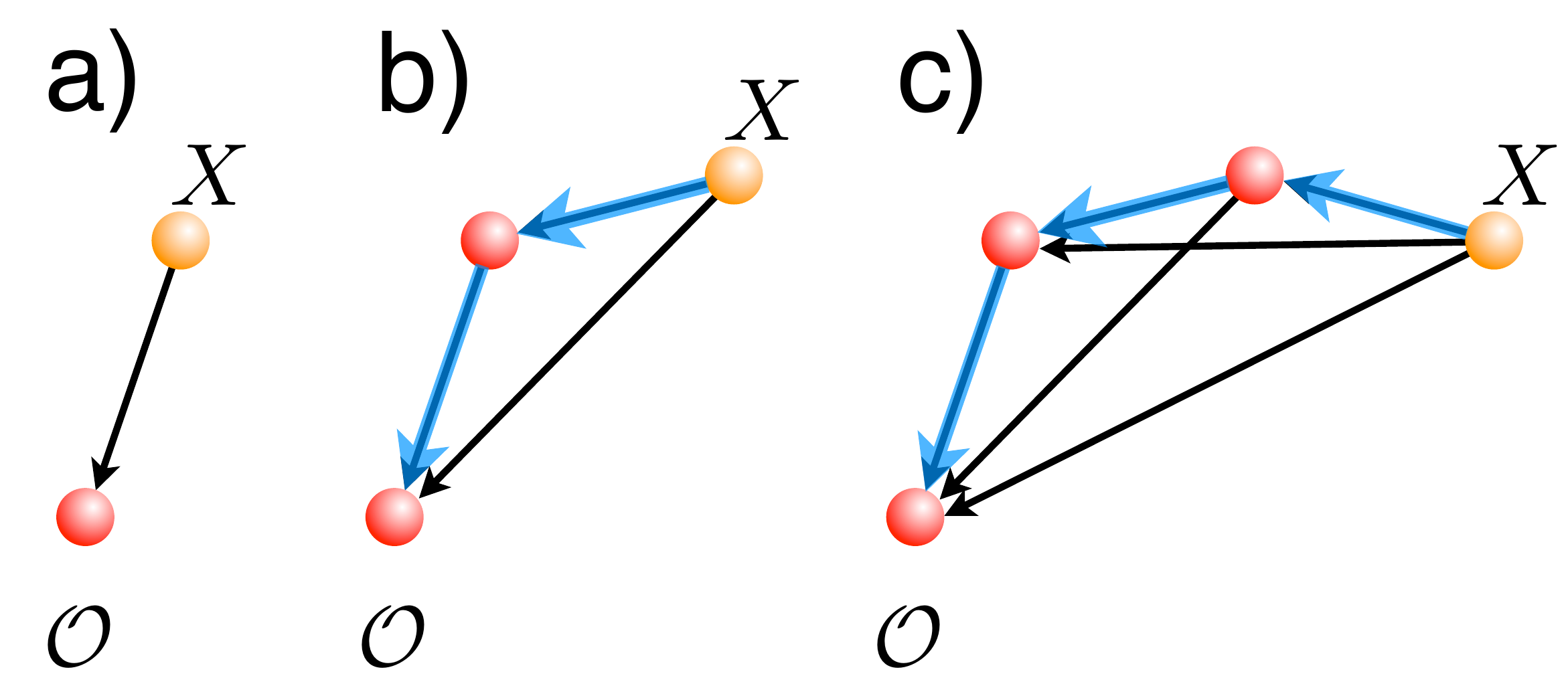}}
\caption{{\bf Direct vs.\ indirect causal relations.} Black edges show the direct causal relations between observers. In b) and c) the blue paths are indirect causal relations between observers~$X$ and~$O$.}
\label{fig:1}
\end{figure}

Here we add to this relatively short list of complex physical networks, a class of
networks that naturally arise in cosmology. Specifically, we consider evolving networks
of causal connections among stationary (co-moving) observers, homogeneously distributed
in any open Friedmann-Lema\^{\i}tre-Robertson-Walker (FLRW)
spacetime~\cite{Weinberg08-book}. These networks are purely classical. Nodes can
represent a dust of classical particles, or (clusters of) galaxies, or indeed imaginary observers,
scattered randomly throughout the space. The horizons of all the observers expand, and
for any particular observer $O$ at any given proper time $\tau$, the network consists of
all other observers within $O$'s horizon, up to a certain cut-off time $\tau_\nu>0$ in
the past, which can be interpreted as the time of last scattering or
the red shift beyond which the observer cannot observe~\cite{Weinberg08-book}.
A directed link from observer $B$ to observer $A$ in this network
exists if $B$ is within $A$'s retarded horizon. The retarded horizon of $A$ is $A$'s horizon
at earlier time $\tau_r<\tau$ such that light emitted by $A$ at time $\tau_r$ reaches $O$ at time $\tau$.
This means that if there are some physical processes running at each observer, then
directed paths between observers~$X$ and~$O$ in this network represent all possible causal relations between~$X$ and~$O$, including indirect relations over paths longer than one hop, Fig.~\ref{fig:1}.
Here we show that this
evolving network of maximum information about the universe that any observer can collect by her
proper time $\tau$, is a growing power-law graph in any open homogeneous and isotropic (FLRW)
spacetime.

We emphasize a critical difference between these cosmological networks and causal sets in de Sitter network cosmology considered in~\cite{KrKi12}. The latter are discretizations of $4$-dimensional spacetime---nodes are elementary events (points in space and time), and two events are connected if they are causally related, i.e., if they lie within each other's light cones. The resulting networks are directed acyclic graphs, and all the linking dynamics is the appearance of new links connecting new nodes to the existing nodes lying in their past light cones. No new links appear between already existing nodes, since any two events are either timelike-separated and thus connected, or spacelike separated and thus disconnected. The cosmological networks considered here are discretizations of $3$-dimensional space. Time remains continuous. Therefore the evolution of nodes in these networks represent world-lines of co-moving observers. These networks have directed cycles, and new links not only connect new nodes to existing ones, but also appear at a certain rate between existing nodes, as they do in many complex networks~\cite{DorMen-book03,newman03c-review,BoLaMoChHw06}.\\

\section{Overlapping horizons in the Milne universe}

The metric in an open FLRW spacetime is given by
\begin{equation}
ds^2=-d\tau^2+R(\tau)^2\left[d\chi^2+\sinh^2{\chi} d \Omega_{d-1}^2\right],
\label{eq:1}
\end{equation}
where $\tau>0$ and $\chi>0$ are the cosmic time and ``radial'' coordinates, $d \Omega_{d-1}^2$ is the
metric on the unit $(d-1)$-dimensional sphere, and $R(\tau)$ is the scale factor of the universe
given by the Friedmann equations~\cite{Weinberg08-book}. The scale factor $R(\tau)$ is just a
conformal factor in the spacial part of the metric, where coordinates $(\chi, \Omega_{d-1})$
describe the hyperbolic $d$-dimensional space $\mathbb{H}^d$ of constant curvature $K=-1$. The
spacetime is thus foliated by $d$-dimensional hyperbolic spaces: for any time $\tau$, the space is
the hyperbolic $d$-dimensional space of constant curvature $K=-1/R(\tau)$. To simplify the
calculations, we assume that $R(\tau)=\tau$, meaning that we are considering the Milne universe---a
completely empty universe without any matter or dark energy~\cite{Mukhanov05-book}. The results
presented henceforth do not depend on a particular form of scale factor $R(\tau)$. We discuss this
important point at the end.

\begin{figure*}[h]
\centerline{\includegraphics[width=\linewidth]{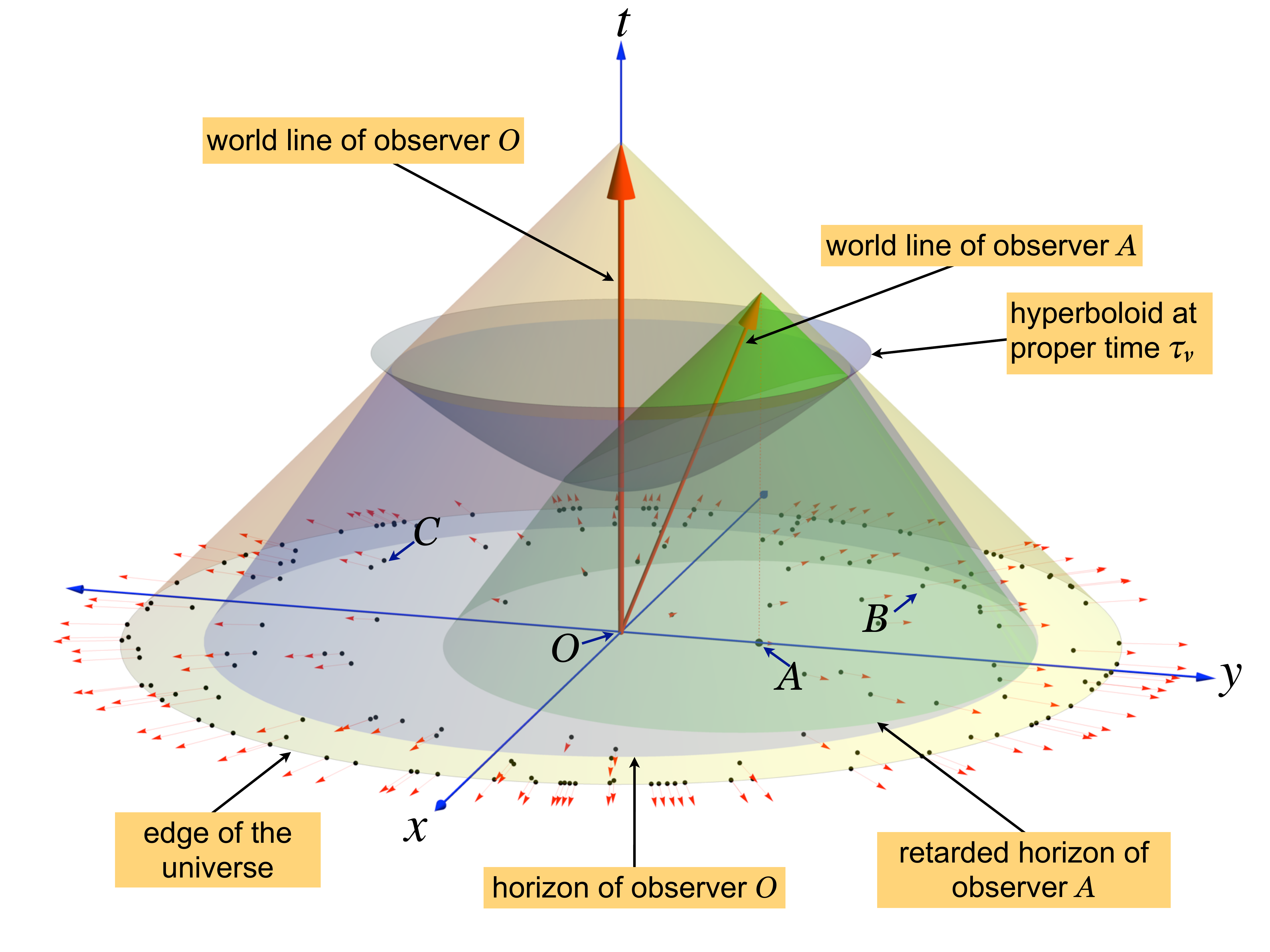}}
\caption{{\bf Milne universe with overlapping horizons as seen by observer $O$ at proper time
$\tau>\tau_{\nu}$}. The horizontal plane is the $(x,y)$ plane in a $(2+1)$-dimensional Minkowski
spacetime, The vertical axis $t$ is the proper time of observer $O$, who is at rest in the
cosmic fluid, $\chi=0$ and $x=y=0$. At cosmic time $\tau=0$, all particles
are at the origin of this Minkowski spacetime and start moving away from $O$ at velocities $v$
according to Eq.~(\ref{speed_distribution}). Points and arrows in the $(x,y)$ plane represent the
position and velocity of such particles at proper time $\tau$ as measured by $O$. The ``edge of
the universe'' corresponds to particles receding from $O$ at the speed of light. This edge is thus a circle of radius $R_{edge}=\tau$ centered at $O$.
Observer $O$ observes not all particles within this edge, since particles are ``lit'' not at $\tau=0$ but at $\tau_{\nu}>0$.
These events lie on the invariant hyperboloid $t^2=\tau_{\nu}^2+x^2+y^2$ shown in blue.
The horizon of any given observer is
then induced by the intersection of her past light cone with this hyperboloid, and defines the
maximum speed of a particle within the horizon. In particular, the radius of $O$'s horizon in the
$(x,y)$ plane is $R_{horizon}=\tau[1-(\tau_{\nu}/\tau)]/[1+(\tau_{\nu}/\tau)]$
($\tau_{\nu}=1.5$ and $\tau=5$ in the figure).
The thick red arrows show the world-lines of
stationary observers $O$ and also $A$ who is at rest at radial coordinate $\chi=\mathrm{const}$.
The retarded horizon of observer $A$ at proper time $\tau_{\chi}$ is induced by the intersection
of $A$'s past light cone with the blue hyperboloid. Projected into the $(x,y)$ plane, this
retarded horizon encompasses all the observers that can causally influence $O$ indirectly via $A$,
Fig.~\ref{fig:1}.
Observer $O$ has incoming connections from observers $A$, $B$, and $C$ since they all lie within
$O$'s horizon. Observer $A$ has incoming connections from $O$ and $B$, but not from $C$ who is
outside $A$'s horizon.}
\label{fig:2}
\end{figure*}

In $(2+1)$ dimensions (the generalization to $(d+1)$ with $d>2$ is straightforward), the change of
coordinates $(\tau,\chi,\theta)$ to
\begin{equation}\label{eq:coordinates}
\begin{array}{ccl}
x&=&\tau \sinh{\chi} \cos{\theta}\\
y&=&\tau \sinh{\chi} \sin{\theta}\\
t&=&\tau \cosh{\chi}
\end{array}
\end{equation}
transforms the metric in Eq.~(\ref{eq:1}) into the Minkowski metric
\begin{equation}
ds^2=-dt^2+dx^2+dy^2.
\label{minkowski}
\end{equation}
However, this transformation does not map the original spacetime in Eq.~(\ref{eq:1}) to the whole
Minkowski spacetime, but only to the future light cone of the event $t=x=y=0$. Indeed, the radial
Minkowski coordinate $r=\sqrt{x^2+y^2}$ of an event at coordinates $(\tau,\chi,\theta)$ is $r=t
\tanh{\chi}$. This means that a stationary observer---that is, an observer at rest in the co-moving
coordinates $(\chi,\theta)$ in $\mathbb{H}^2$---is receding from the origin $x=y=0$ at constant
speed $v=\tanh{\chi}\le 1$. Consistent with homogeneity and isotropy of the universe, we assume
that stationary observers are also homogeneously and isotropically distributed throughout space
with constant density $\delta$. These observers are therefore points distributed in the hyperbolic
space $\mathbb{H}^2$ according to a Poisson point process with point density $\delta$. In the Milne
cosmology, an infinite number of such observers are thus initially at the origin of coordinates (the big bang),
and then they all start moving in all directions within a bubble --in the considered case this bubble
is a disk in $\mathbb{R}^2$-- that expands at the speed of light (see the $(x,y)$ plane in~Fig.~\ref{fig:2}).
Because the distribution of observers is uniform in
$\mathbb{H}^2$, any stationary observer will ``see'' all other observers receding from her with the
Lorentz-invariant density of speeds $v$
\begin{equation}
\delta(v) \propto \delta \frac{v}{(1-v^2)^{3/2}}.
\label{speed_distribution}
\end{equation}

Without loss of generality or breaking Lorentz invariance, in what follows we focus on the
stationary observer $O$ at rest at coordinate $\chi=0$, and therefore also at rest at
$x=y=0$. According to Eq.~(\ref{eq:coordinates}), $O$'s proper time $\tau$ is equal to
the time coordinate $t$ in the Minkowski spacetime. First, we determine the horizon of
$O$ at any given proper time $\tau$. This horizon is the radius of the part of the
universe that $O$ can observe, up to the past cut-off time $\tau_\nu$, which can be any
positive number, $0<\tau_\nu<\tau$.
This radius is determined by the intersection of $O$'s past light cone with the hyperboloid at time $\tau_\nu$, Fig.~\ref{fig:2}. At time $\tau > \tau_{\nu}$, the farthest particle that $O$ can observe is moving at a speed such that light emitted at proper time $\tau_{\nu}$ reaches $O$ at this time $\tau$, yielding the following simple expression for the hyperbolic radius of $O$'s horizon:
\begin{equation}\label{eq:horizon}
\chi_h=\ln{\left(\frac{\tau}{\tau_{\nu}}\right)}.
\end{equation}
The size of the network, i.e., the number is nodes in it, is in this case the number of other observers that $O$ can observe, equal to the number of points within a
hyperbolic disk of radius $\chi_h$. This number grows asymptotically linearly with time $\tau$:
\begin{equation}
N(\tau)=2 \pi \delta (\cosh{\chi_h}-1)=\pi \delta \left[ \frac{\tau}{\tau_{\nu}}+\frac{\tau_{\nu}}{\tau}-2\right] \approx \pi \delta \frac{\tau}{\tau_{\nu}}.
\label{N_tau}
\end{equation}
Any two observers $A$ and $B$ in $O$'s horizon are connected by a directed link from $B$ to $A$ if $B$
lies within the retarded horizon of $A$. If $A$'s radial coordinate is~$\chi$, then the retarded horizon of $A$ is defined as its horizon at time $\tau_\chi=\tau e^{-\chi}$.
According to Eq.~(\ref{eq:horizon}), $\tau_\chi$ is such that if $A$ emits light at her proper time $\tau_\chi$, then this light reaches
$O$ at time $\tau$. This means that if $A$ has some physical state (possibly causally influenced by $B$) at time $\tau_\chi$, then this
state can causally influence $O$ by time $\tau$.

Figure~\ref{fig:2} shows observer $A$ lying within the horizon of observer $O$.
Observer $B$ is connected to $A$ because $B$ lies within $A$'s retarded horizon at time
$\tau_\chi$, the latest time in $A$'s history that can influence $O$ at time $\tau$.
Observer $C$ is outside of this horizon and therefore is not connected to $A$. The link between
$O$ and $A$ is bi-directed because they lie within each other horizons.

\begin{figure}[t]
\centerline{\includegraphics[width=0.8\linewidth]{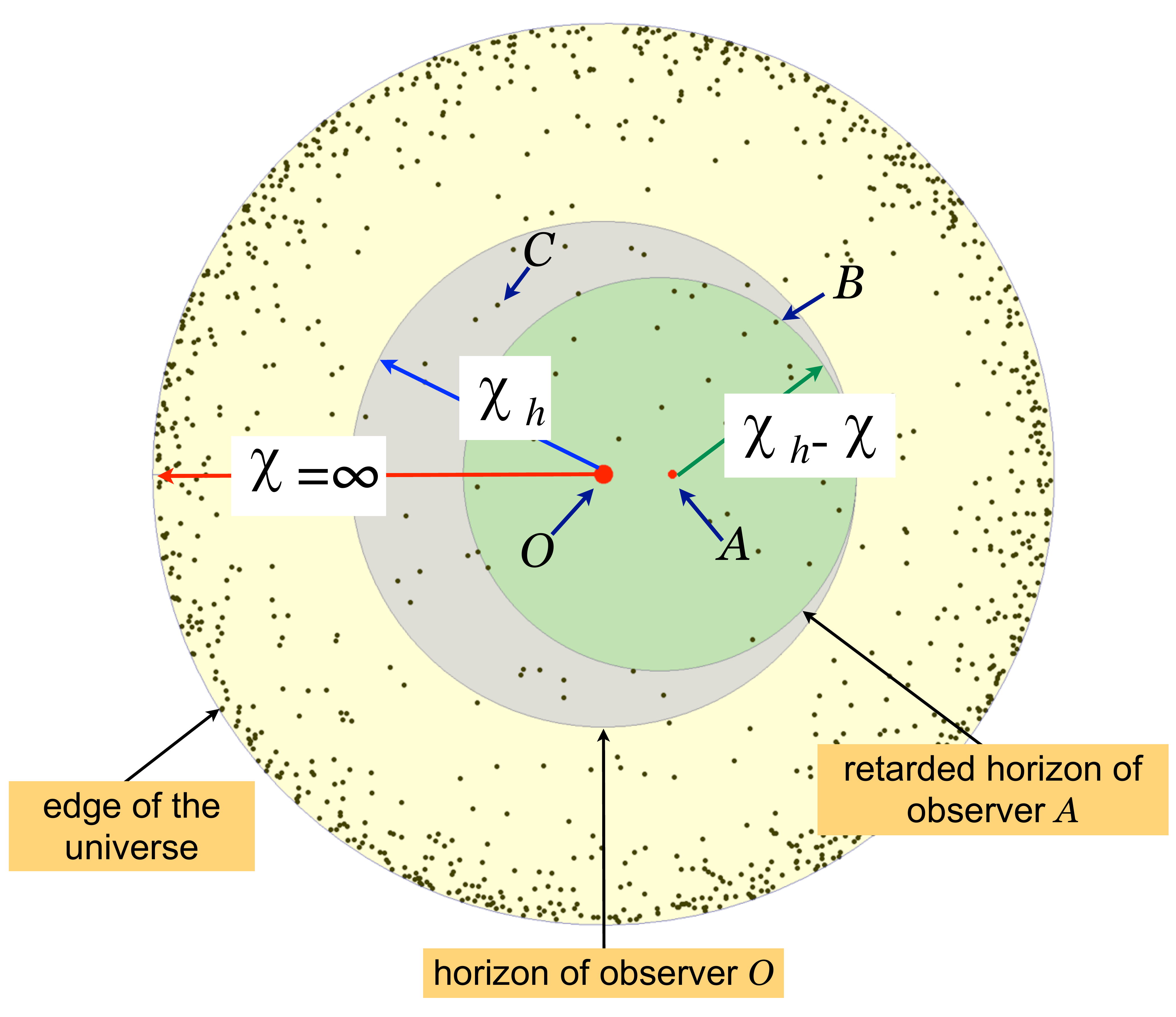}}
\caption{{\bf Milne universe projected onto the hyperbolic plane.} All moving observers in
Fig.~\ref{fig:2} and their horizons can be mapped to the hyperbolic plane $\mathbb{H}^2$ via the change of
coordinates in Eq.~(\ref{eq:coordinates}). After the mapping, observers become static points on
$\mathbb{H}^2$, while their horizons expand with cosmic time. The blue
area is the horizon of observer $O$ of hyperbolic radius $\chi_h$. The green area
is $A$'s retarded horizon of radius $\chi_h-\chi$, centered at $A$ and
tangent to $O$'s horizon. Nodes $B$ and $C$ are the same as in Fig.~\ref{fig:2}.
The picture does not depend on scale factor $R(\tau)$, which determines only how
horizon $\chi_h$ grows with cosmic time $\tau$.
}
\label{fig:3}
\end{figure}

Mapped to the hyperbolic plane, the horizon of observer $O$ is a disk
of radius $\chi_h$, whereas the horizon of observer $A$ is a disk of radius $\chi_h-\chi$ centered
at $A$ who is located at radial coordinate $\chi$. This disk is tangent to $O$'s horizon
as illustrated in Fig.~\ref{fig:3}. The expected number of direct incoming connections to observer $A$, i.e., $A$'s in-degree $\bar{k}_{in}(\chi)$,
is thus given by the number of points within a disk of radius $\chi_h-\chi$:
\begin{equation}
\bar{k}_{in}(\chi)=2 \pi \delta (\cosh{(\chi_h-\chi)}-1) \approx \pi \delta e^{-(\chi-\chi_h)}.
\end{equation}
On the other hand, since observers are distributed uniformly according to the hyperbolic metric,
the density of them located at radial coordinate $\chi$ is given by distribution
\begin{equation}
\rho(\chi)=\frac{\sinh{\chi}}{\cosh{\chi_h}-1} \approx e^{\chi-\chi_h}.
\label{rho_chi}
\end{equation}
We thus have a combination of two exponential dependencies: $\bar{k}_{in}(\chi)\sim e^{-\chi}$ and $\rho(\chi)\sim e^\chi$. As one can check~\cite{newman05}, if in general the expected value $\bar{k}(x)$ of some random variable $k$ decays exponentially, $\bar{k}(x) \sim e^{-\alpha x}$, $\alpha>0$, as a function of random variable $x$ whose distribution is also exponential, $\rho(x) \sim e^{\beta x}$, $\beta>0$, then the distribution of $k$ is a power law, $P(k) \sim k^{-\gamma}$, with exponent $\gamma = \beta/\alpha+1$. In our case $\alpha=\beta=1$, so that $\gamma=2$:
\begin{equation}
P(k_{in}) \sim \frac{1}{k_{in}^2},\quad\text{if $1\ll k_{in} < \pi \delta e^{\chi_h}$}.
\end{equation}
In large networks with $\chi_h \gg1$ the average in-degree scales as $\langle
k_{in}\rangle \sim \pi \delta \chi_h \approx \pi \delta \ln{(N/\pi \delta)}$.
The degree distributions in many large real networks are also close to power laws with exponents close to $2$~\cite{DorMen-book03,newman03c-review,BoLaMoChHw06}.

Next we focus on the expected number of out-going connection, i.e., out-degree, of a node located at $(\chi,0)$. It is equal to the number of points within a domain in $\mathbb{H}^2$
defined as the locus of points $(\chi',\theta)$ such that their hyperbolic distances to the point $(\chi,0)$, $x$, are smaller than
the radius of their retarded horizons $\chi_h-\chi'$, that is,
\begin{equation}
\bar{k}_{out}(\chi)=2  \delta \int_{0}^{\pi}d\theta \int_0^{\chi_h} d \chi'\,\sinh{\chi'}\,\Theta(\chi_h-\chi'-x),
\end{equation}
where $\Theta(\cdot)$ is the Heaviside step function.
In the limit $\chi_h\gg 1$, the integration yields
\begin{equation}\label{eq:out-degree}
\bar{k}_{out}(\chi) \approx
\begin{cases}
 2\delta \sqrt{\displaystyle{\frac{e^{\chi_h}}{\cosh{\chi}}}}\,K\left(\tanh{\chi}\right) &\text{if $0 \le \chi<\chi_h$,}\\
 0 &\text{if $\chi=\chi_h$,}
\end{cases}
\end{equation}
where $K(\cdot)$ is the complete elliptic integral of the first kind. In the regime
$1<\chi<\chi_h$, the average out degree is well approximated by
\begin{equation}
\bar{k}_{out}(\chi) \approx 2 \sqrt{2} \delta \chi e^{(\chi_h-\chi)/2}.
\end{equation}
For the same combination-of-exponentials reasons as in the in-degree case, this exponential scaling, combined with the one in Eq.~(\ref{rho_chi}), implies  that the out-degree distribution scales as
\begin{equation}
P(k_{out}) \sim k_{out}^{-3},\quad\text{for $k_{out}\gg1$,}
\end{equation}
with logarithmic corrections. We notice however that observers near (but not exactly at) the edge
of the horizon have out-degrees approximately equal
to $\chi_h$. Therefore, the out-degree distribution is asymptotically a power law with a lower cut-off that grows as
$\chi_h$ with time.

\begin{figure*}[t]
\centerline{\includegraphics[width=\linewidth]{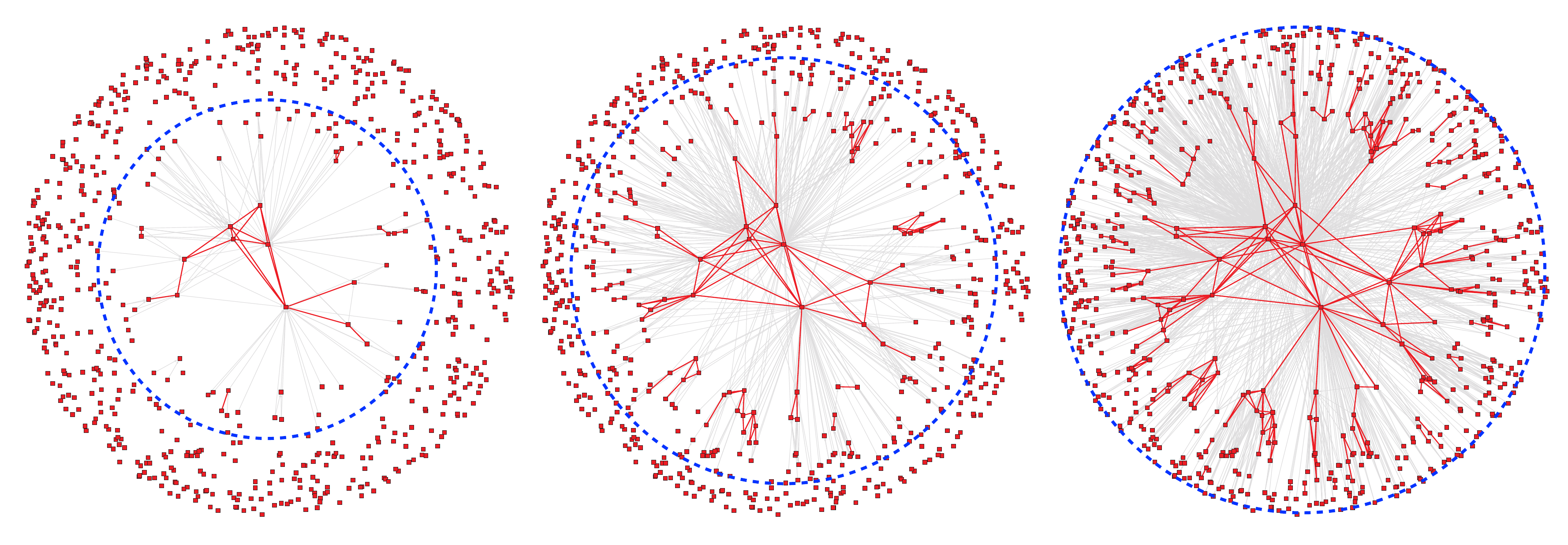}}
\caption{{\bf Evolution of a Milne network at three
different proper times.} The dashed blue circles represent the expanding horizons of the central
observer. The grey and red links show directed and bi-directed (reciprocal) connections. The
central observer and all her connections are suppressed.}
\label{fig:4}
\end{figure*}

We note that new connections appear not only between new and existing nodes, but also between pairs of already existing nodes, not previously connected. This type of linking creates directed cycles in the network. The appearance of new links between existing nodes is a simple consequence of the continuous expansion of the horizons of all observers. The resulting network dynamics is illustrated in Fig.~\ref{fig:4}, where three snapshots of a growing network are taken. The horizon of the central observer $O$
(the blue dashed circle) grows over time, discovering an exponentially increasing number of new
observers. Gray connections indicate purely directed causal relations between observers, that is,
one is aware of the other. As time goes on, directed connections are reciprocated (connections in
red), meaning that an increasing number of pairs of observers are getting mutually aware of each
other.

Finally we emphasize that our analysis is by no means limited to the Milne universe.
Almost exactly the same results hold for any open FLRW universe with any scale factor $R(\tau)$. The same picture as in
Fig.~\ref{fig:3} would apply there. The only minor difference is the rate at which new nodes join the
network, defined by the radius of the observer's horizon as a function of time. Specifically, given $R(\tau)$, this
radius is
\begin{equation}
\chi_h= \int^{t} \frac{d \tau}{R(\tau)},
\end{equation}
generalizing Eq.~(\ref{eq:horizon}).
\\

\section{Imperfect communication}

\begin{figure}[t]
\centerline{\includegraphics[width=0.8\linewidth]{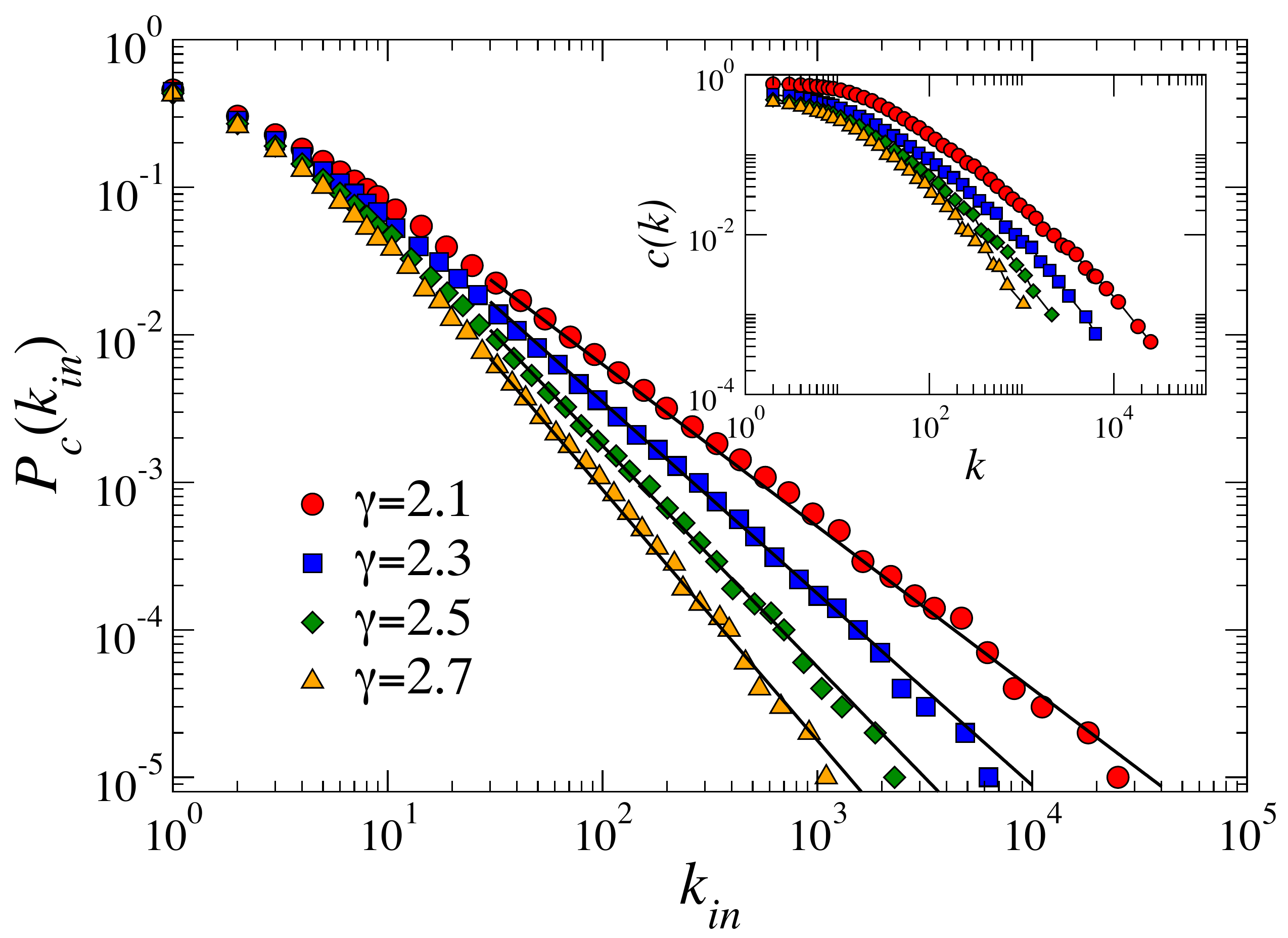}}
\caption{Complementary cumulative in-degree distribution $P_c(k_{in})=\sum_{k_{in}' \geq k_{in}}P(k_{in}')$ in simulated Milne networks with exponents $\gamma=2.1,2.3,2.5,2.7$ grown up to $N=10^5$ nodes. The solid lines are power laws with the same exponents. Inset: degree-dependent clustering coefficient
for the undirected versions of the same networks. The average clustering coefficients excluding nodes of degree~$1$ are $\bar{C}=0.67, 0.47, 0.41, 0.38$ for $\gamma=2.1,2.3,2.5,2.7$, respectively. The networks are disassortative, meaning that the correlations of degrees of connected nodes (not shown) are negative, due to structural constraints imposed by the scale-free degree distribution~\cite{Boguna:2004eh}.
}
\label{fig:5}
\end{figure}

Up to this point we have assumed that all observers entering the horizon of another observer are
detected with probability $1$. If we assume that the probability of connection between observers decays exponentially with the
hyperbolic distance $x$ between them,
\begin{equation}\label{eq:imperfect-p(x)}
p(x) =p e^{-\beta x},
\end{equation}
then the average in-degree of an observer at coordinate $\chi$ is
\begin{eqnarray}
\bar{k}_{in}(\chi)&=&2\pi \delta p \int_0^{\chi_h-\chi} \sinh{\chi'}\,e^{-\beta \chi'} d \chi'\nonumber\\
&=&2\pi \delta p\frac{1-e^{\beta(\chi-\chi_h)}[\beta \sinh{(\chi_h-\chi)}+\cosh{(\chi_h-\chi)}] }{\beta^2-1}.
\end{eqnarray}
If $\beta \ge 1$ and $\chi_h\gg 1$ the average in-degree of nodes is constant and the network becomes similar to a random geometric graph. In random geometric graphs, nodes lie in a geometric space, and two nodes are connected if the distance between them in the space is below a given threshold. These graphs have Poisson distributions of node degrees~\cite{Penrose03-book}. We can show that the in-degree distribution in our imperfect networks with $\beta\geq1$ is also Poisson. This is intuitively expected because in this case observers are connected only to other observers in their small neighborhoods. The case with $\beta<1$ is more interesting. In this case, the average in-degree of an observer located at $\chi$ is $\bar{k}_{in}(\chi) \sim e^{(1-\beta)(\chi_h-\chi)}$. As a consequence, for the same combination-of-exponentials reasons as before, the in-degree distribution scales asymptotically as a power law $P(k_{in}) \sim k_{in}^{-\gamma}$ with exponent
\begin{equation}
\gamma=2+\frac{\beta}{1-\beta},
\end{equation}
which can take any value between $2$ and $\infty$, as shown in Fig.~\ref{fig:5}.

This result may have interesting cosmological implications concerning what part of the universe our observers can observe.
Indeed, in the case of imperfect communication with $\beta\in(0,1)$, observer $O$ directly detects only $\sim e^{(1-\beta)\chi_h}$ other observers. Therefore by the time the number of observers within $O$'s horizon is $\sim N$, $O$ detects only $\sim N^{1-\beta}$ of those, so that the fraction of the universe that $O$ sees directly ($\sim N^{1-\beta}/N$) approaches zero as time goes on. However there are also indirect causal paths, shown in blue in Fig.~\ref{fig:1}. Any observer connected to $O$ via either direct or indirect causal paths can still be detected by $O$. The question of what fraction of the universe can be observed by $O$ becomes a variation of the bond percolation problem, well studied in network science. In the classical bond percolation problem, we are given a large network in which we retain or delete each link (also called ``bond'' for historical reasons) with probability~$p$ and $1-p$. There often exists a critical value $p_c$ of this probability corresponding to the phase transition in the system: if $p>p_c$ the network is in the percolated phase, meaning that a macroscopical fraction of nodes belong to the largest connected component, while for $p<p_c$ the network decomposes into many small connected components. There is no such phase transition in random scale-free networks with power-law exponent $\gamma<3$. They are always in the percolated phase, $p_c=0$~\cite{SeKrBo11}.
In our imperfect cosmological networks with $\beta\in(0,1)$, the given network is the perfect network with $\beta=0$ and $p=1$ in which we retain links with probability in Eq.~(\ref{eq:imperfect-p(x)}), and the question is now what fraction of the network is connected to $O$ via at least one causal path, direct or indirect. This problem is more involved that the standard bond percolation problem, but one may suspect that since the network is scale-free, there should exist a regime, perhaps with $\beta<1/2$, in which the network is always percolated. This would imply that a macroscopic fraction of the universe can be observed by any observer.
\begin{figure}[t]
\centerline{\includegraphics[width=0.8\linewidth]{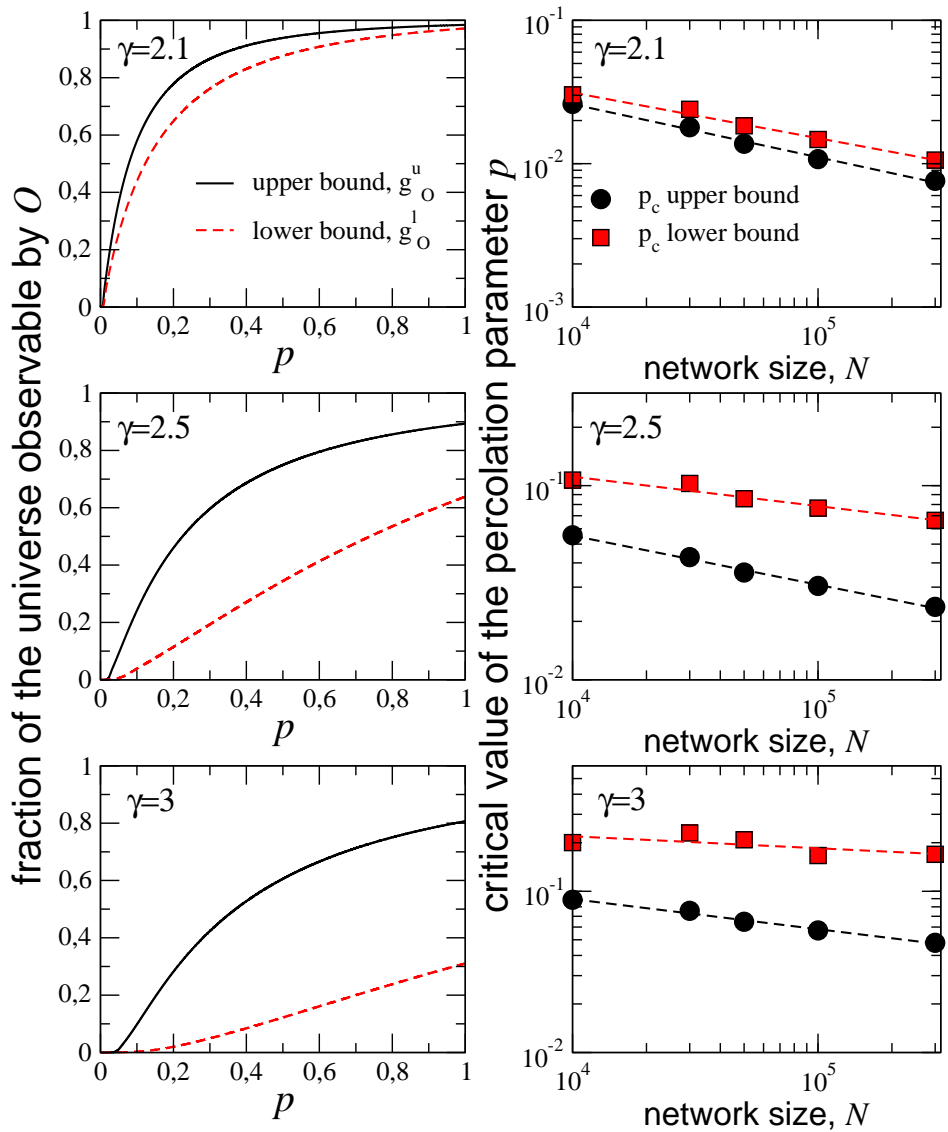}}
\caption{Bond percolation simulations of imperfect-communication networks. The left column shows the upper and lower bounds $g_O^{u,l}(p)$ for the fraction of nodes causally connected to $O$ for different values of $\gamma$ and $N=3 \times 10^5$. The right column shows the critical values $p_c^{u,l}(N)$ for the same bounds, measured as the value of $p$ that maximizes the susceptibilities $\xi^{u,l}$ in Eq.~(\ref{chi}). The dashed lines are power law fits $p_c^{u,l}(N) \sim N^{-1/\nu}$ with exponents $1/\nu=0.3(7), 0.3(2)$ for $\gamma=2.1$, $1/\nu=0.2(5), 0.1(5)$ for $\gamma=2.5$, and $1/\nu=0.1(8), 0.0(7)$ for $\gamma=3$.}
\label{fig:6}
\end{figure}

We next support these expectations in simulations. Let $g_{O}$ be the fraction of nodes within $O$'s horizon that are connected to $O$ via at least one causal path in an imperfect-communication network with the link existence probability~(\ref{eq:imperfect-p(x)}). From the exposition above, including Fig.~\ref{fig:1}, the causal path is defined as a directed path $P=\{n_1,n_2,\ldots,O\}$ such that the retarded horizon $H_{n_i}$ of any node $n_i$ in the path, $i=1,2,\ldots$ (or equivalently the set of $n_i$'s neighbors in the perfect network), contains all subsequent nodes in the path: $n_j \in H_{n_i}$ for any $j>i$. The problem of finding if such a path exists between a given node $n_1$ and $O$ is likely to be an NP-hard combinatorial problem, because checking all directed paths between $n1$ and $O$ seems unavoidable. We did not attempt either to prove the NP hardness of the problem or to find its computationally admissible solution, because it is much easier to provide upper and lower bounds for $g_{O}$. An upper bound $g_{O}^u$ is just the number of nodes connected to $O$ via any directed path, not necessarily causal. As a lower bound $g_{O}^l$ we use the number of nodes connected to $O$ by at least one causal path, and located up to three hops away from $O$, which comprise a significant fraction of all nodes within $O$'s horizon.
\begin{figure}[t]
\centerline{\includegraphics[width=0.8\linewidth]{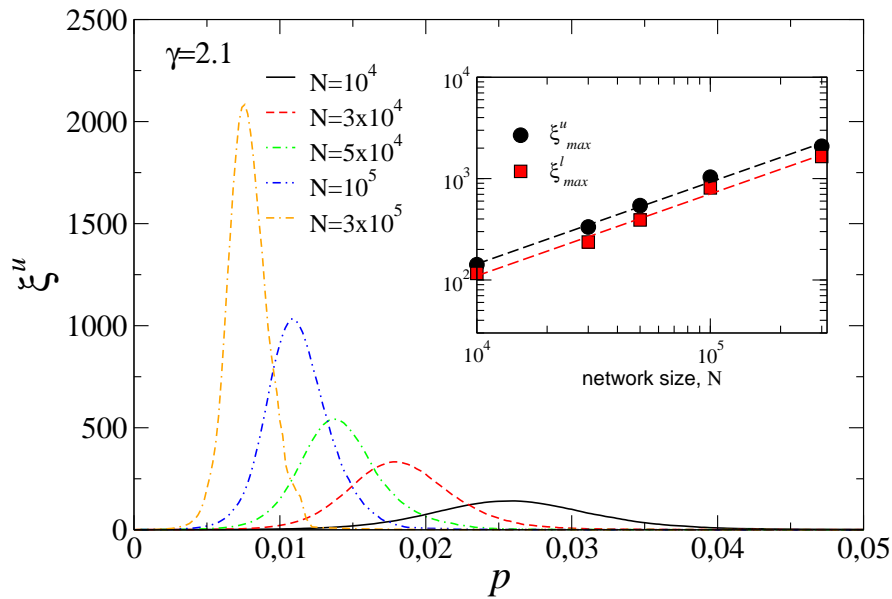}}
\caption{Percolation susceptibility of the upper bound $\xi^u$ Eq.~(\ref{chi}) as a function of $p$ for different network sizes and $\gamma=2.1$ (results for other values of $\gamma$ are qualitatively similar). For all values of $\gamma<3$, $\xi^u$ and $\xi^l$ show a peak that moves to the left as the system size increases. At the same time, the maximum value of $\xi^u$ and $\xi^l$ diverges as a function of $N$ as $\xi^{u,l}_{max}\sim N^{\gamma'/\nu}$. Dashed lines in the inset are power law fits with exponents $\gamma'/\nu=0.8(0)$.
\label{fig:7}}
\end{figure}

Figure~\ref{fig:6} shows the results for these bounds in numerical simulations of networks with up to $N=3 \times 10^5$ nodes, $\gamma=2.1,2.5,3$ and $p\in[0,1]$. The upper and lower bounds $g_{O}^u$ and $g_{O}^l$ increase monotonically as functions of $p$, suggesting that, as expected, the percolation threshold is zero. To check if it is indeed zero, we measure the susceptibilities $\xi^u$ and $\xi^l$ defined as
\begin{equation}
\xi^{u,l}=N\frac{\langle [g_{O}^{u,l}]^2 \rangle -\langle g_{O}^{u,l} \rangle^2}{\langle g_{O}^{u,l} \rangle},
\label{chi}
\end{equation}
where averages $\langle\cdot\rangle$ are taken over a large number (10000 in our case) of different bond percolation realizations for each combination of values of $N$, $\gamma$, and $p$. In continuous phase transitions, the fluctuations of a property of interest ($\xi$ in our case) diverge at a critical parameter value in the thermodynamic limit $N\to\infty$. In finite-size systems, this divergence manifests itself as a maximum of function $\xi(p)$ that becomes sharper for larger $N$, see Fig.~\ref{fig:7}. The value of $p=p_c$ corresponding to this maximum can be used as an estimate of the critical parameter value $p_c$~\cite{Newman99-book}. The right column in Fig.~\ref{fig:6} shows the values of thus estimated $p_c$s as functions of $N$ for bounds $\xi^{u,l}(p)$ in our networks. For $\gamma<3$ the critical points of both upper and lower bounds go to zero as power laws $p_c \sim N^{-1/\nu}$. This means that the percolation threshold is indeed zero in the thermodynamic limit ($p_c\to0$ as $N\to\infty$), and that observer $O$ can observe a finite fraction of the universe for any value of $p$. However if $\gamma=3$, then while the upper bound critical value goes to zero as $N$ goes to infinity, the critical value corresponding to the lower bound becomes nearly size independent. This implies that for $\gamma>3$, there exists a critical point $p_c$ below which our observer $O$ can observe only her local neighborhood.\\

\section{Conclusions}

In summary, the physical network of (indirect) causal relations between observers
uniformly distributed in any open FLRW universe is a Lorentz-invariant scale-free graph with strong clustering, Fig.~\ref{fig:5}.
This network represents maximum information about the
universe that any particular observer can collect by a certain time. More precisely, paths in this
network are all possible communication channels between observers. Perhaps coincidentally, in the
perfect case without information loss ($\beta=0$), this network has the same statistical properties
($\gamma=2$ and strong clustering) as the maximally navigable networks~\cite{BoKrKc08}, i.e., networks
that are most conductive with respect to targeted information signaling.
The crucial requirement for this coincidence is that the universe must be open, Eq.~(\ref{eq:1}).
Bubble universes are open in most inflationary cosmologies~\cite{Kleban2011}, and the current
measurements of our universe do not preclude that it is open either, although it is definitely
close to being flat~\cite{Komatsu11}.

These results may be interesting for both network science and cosmology. From the network science perspective, they may help to develop a ``General Relativity'' of networks, an analogy of the Einstein equations that would describe network dynamics within a unified framework, in which network nodes might be analogous to our observers or galaxies.
Here we have considered an idealized case where nodes are
massless points distributed uniformly in the space. It remains unclear how the picture would
change if points have masses, perhaps distributed according to some heterogeneous distributions
similar to the distribution of the masses of galaxies in the universe~\cite{FoPo04},
and if the spatial distribution of points deviates from uniform, as it does for
galaxies~\cite{LaPe010} and for real networks embedded in hyperbolic
spaces~\cite{PaBoKr11}.

From the cosmology perspective, it has been suggested that measures of photons from the cosmic microwave background scattered by high energy electrons in clusters of galaxies could be used to probe the last scattering surface (LSS) at many different length scales, and thus overcome the limitations of the cosmic variance~\cite{Kamionkowski:1997fk}. In this context the cosmological networks we have considered here may be interesting because they contain not only direct connections within causal horizons, but also all possible indirect causal connections. The galaxy-scattered photons represent the latter indirect connections between the LSS and us, albeit made of only two hops, as in Fig.~\ref{fig:1}~b. Yet the knowledge of the density of clusters of galaxies throughout the universe, coupled with our network representation, can be used to estimate the maximum information we could ever obtain from the LSS by counting the total number of causal paths connecting such surface to us. The discussed percolation problem on these networks may be of particular interest in that respect.\\

\ack
We thank Jaume Garriga for very useful comments and suggestions. This work was supported by a James S.\ McDonnell Foundation Scholar Award in Complex Systems; the ICREA Academia prize, funded by the {\it Generalitat de Catalunya}; MICINN project No.\ FIS2010-21781-C02-02; {\it Generalitat de Catalunya} grant No.\ 2014SGR608; DARPA grant No.\
HR0011-12-1-0012; NSF grants No.\ CNS-1344289, CNS-0964236, and CNS-1039646; and by Cisco Systems.\\

%\bibliographystyle{unsrt.bst}
%\bibliography{bib}

\end{document}